\documentclass[aps,showpacs,two column]{revtex4}

\usepackage{graphicx}
\usepackage[squaren]{SIunits}

\usepackage{amssymb}
\usepackage{color}
\begin{document}
\title{SDW transition of Fe1 zigzag chains and metamagnetic transition of Fe2 in TaFe$_{1+y}$Te$_3$ }

\author{R. H. Liu, M. Zhang, P. Cheng, Y. J. Yan, Z. J. Xiang, J. J. Ying, X. F. Wang, A. F. Wang, G. J. Ye, X. G. Luo and X. H. Chen}
\altaffiliation{Corresponding author} \email{chenxh@ustc.edu.cn}
\affiliation{Hefei  National Laboratory for Physical Science at
Microscale and Department of Physics, University of Science and
Technology of China, Hefei, Anhui 230026, People's Republic of
China}

\begin{abstract}

We systematically study the AFM order of Fe1 zigzag chains and
spin-flop of excess Fe2 under high magnetic field H through the
susceptibility, magnetoresistance (MR), Hall effect and specific
heat measurements in high-quality single crystal TaFe$_{1+y}$Te$_3$.
These properties suggest that the high temperature AFM transition of
the TaFeTe$_3$ layers should be a SDW-type AFM order. Below T$_N$,
Fe1 antiferromangetic zigzag chains will induce a inner magnetic
field \textbf{H$_{int}$} to interstitial Fe2 and lead Fe2 also forms
an AFM alignment, in which the magnetic coupling strength between
Fe1 and Fe2 is enhanced by decreasing temperature. On the other
hand, the external magnetic field \textbf{H$_{ext}$} inclines to
tune interstitial Fe2 to form FM alignment along \textbf{H$_{ext}$}.
When \textbf{H$_{ext}$} arrives at the `` coercive" field H$_C$,
which is able to break the coupling between Fe1 and Fe2, these
interstitial Fe2 atoms take a spin-flop from AFM to FM alignment.
The local moment of Fe2 is about 4 $\mu_{\textrm{B}}$/Fe. From low
field ($<$H$_C$) AFM to high field ($>$H$_C$) FM for Fe2, it also
induces sharp drop on resistivity and an anomalous Hall effect. The
possible magnetic structure of TaFe$_{1+y}$Te$_3$ is proposed from
the susceptibility and MR. The properties related to the spin-flop
of Fe2 supply a good opportunity to study the coupling between Fe1
and Fe2 in these TaFe$_{1+y}$Te$_3$ or Fe$_{1+y}$Te with
interstitial Fe2 compounds.

\end{abstract}
\vskip 15 pt 
\maketitle

\section{Introduction}
The discovery of iron-based high temperature superconductors has
generated great interests in exploring layered Fe-based pnictides
and chalcogenides\cite{Kamihara, xhchen, ren3, hsu}. The
superconducting transition temperature (T$_c$) in iron chalcogenides
$\alpha$-FeSe increased from initial 8 K \cite{hsu} to 15 K by
partial Te substitution for Se \cite{yeh, fang}, and up to 37 K
under high pressure\cite{mizuguchi, medvedev}. Recently, the new
intercalated iron selenides $A_x$Fe$_{2-y}$Se$_2$ ($A$ = K, Rb, Cs
and Tl) \cite{xlchen, fang1, Mizuguchi, Wang} were reported to have
T$_c$ around 32 K, even 43 K. Although the presently known maximum
critical temperatures are lower than that of the iron pnictides,
these iron chalcogenides have attracted considerable attention due
to the fact that it is virulent As free, and that they have very
interesting coexistence and competition relation between magnetism
and superconductivity\cite{Bendele, Margadonna}. The observation of
spin resonance below T$_c$ and enhancement of spin fluctuation near
T$_c$ in iron chalcogenides suggest a superconducting pairing
mechanism mediated by spin fluctuation. The high pressure and muon
spin rotation ($\mu$SR) experiments indicate the static magnetic
phase microscopically coexists with superconductivity in
FeSe$_{1-x}$ under pressure. In addition, the magnetic order
temperature (T$_N$) and superconducting transition temperature T$_c$
are both enhanced by pressure\cite{Bendele, Margadonna}. In the
intercalated iron selenides, $\mu$SR\cite{sher, Vyu}, neutron
scattering\cite{BaoWW} and high temperature magnetization and
resistivity\cite{lrh} also indicate that superconductivity coexists
with antiferromagnetic (AFM) with high T$_N$ (T$_N$=470-550 K) and
large magnetic moment 2-3.3 $\mu_{\textrm{B}}$/Fe\cite{BaoWW, Vyu}.
Furthermore, Fe$_{1+y}$Te has the most simple crystal structure in
iron based superconductors. It is stacked with anti-PbO-type FeTe
layer along c-axis, in which iron atoms (Fe1) form a square plane in
the edge-sharing FeTe tetrahedral layer. Fe$_{1+y}$Te always
contains excess iron atoms (Fe2), which randomly occupy the
interstitial sites of FeTe layer and directly couple with the four
nearest neighbor Fe1 atoms in the iron square plane\cite{fredrik,
fruchart}. This structural characteristic is analogous to that of
PbFCl-type Fe$_2$As, where one half iron atoms (Fe1) and As form
edge-sharing tetrahedral network, and the other half of iron atoms
(Fe2) fully occupy these interstitial sites between anti-PbO type
FeAs layers. Fe$_2$As compound also has a SDW-type AFM order at 353
K\cite{hisao}, which is higher than T$_{SDW}$ in FeAs-based parent
compounds.

    The Fe$_{1+y}$Te is not superconducting until Te atoms
are partially replaced by Se or S and excess Fe atoms (Fe2) are also
removed simultaneously\cite{yeh, fang}. Similar to iron-pnictides,
Fe$_{1+y}$Te exhibits a structural and AFM transition simultaneously
near T$_N$$\sim$ 60-70 K. The different T$_N$ arise from the
different contents of excess partial iron atoms (Fe2)\cite{fredrik,
fruchart, hsu, yeh, fang}. However, its AFM structure is distinct
from that in FeAs-based parent compounds. A collinear commensurate
AFM order with Fe moment along a-axis has been identified in the
iron-pnictides\cite{ccruz, huang}, while Fe$_{1+y}$Te has a
bicollinear and 45$^o$ rotated AFM order\cite{Bao4, li}. In
addition, neutron scattering experiment found that interstitial Fe2
could tune the AFM wave vector from commensurate to incommensurate
in Fe$_{1+y}$Te when $y$ is increased to above 0.076. Theory
suggests that the interstitial Fe2 with a valence near Fe$^+$
donates charge to the FeTe layers\cite{zhang}. There is also a very
strong tendency toward moment formation on the Fe2, then these
interstitial Fe2 with a large local moment will interact with the
magnetism of the FeTe layers, complicating the magnetic
order\cite{Bao4, li, zhang, liutj}. Here we report a
mixed-metal-network layered compound TaFe$_{1+y}$Te$_3$, which has
also partial iron atoms Fe2 randomly occupying these interstitial
sites of tetrahedral (Ta,Fe)Te\cite{badding, liusx, perez}, similar
to that in Fe$_{1+y}$Te. TaFe$_{1+y}$Te$_3$ has a SDW-type AFM
transition at 160 - 200 K depending on the contents of interstitial
Fe2 atoms. In addition, below T$_N$ these Fe2 atoms directly couple
with the three nearest Fe1 atoms of Ta-Fe1 mixed network layers and
also form AFM alignment, but external magnetic field (H$_{ext}$) is
able to break the coupling between Fe1 and Fe2, and cause Fe2 to
take a spin-flop and form FM alignment from AFM alignment. It offers
an interesting opportunity to investigate the interplay between
excess Fe2 and magnetism and transport properties of Ta-Fe mixed
network layer.

\begin{figure}[h]
\includegraphics[width=0.50\textwidth]{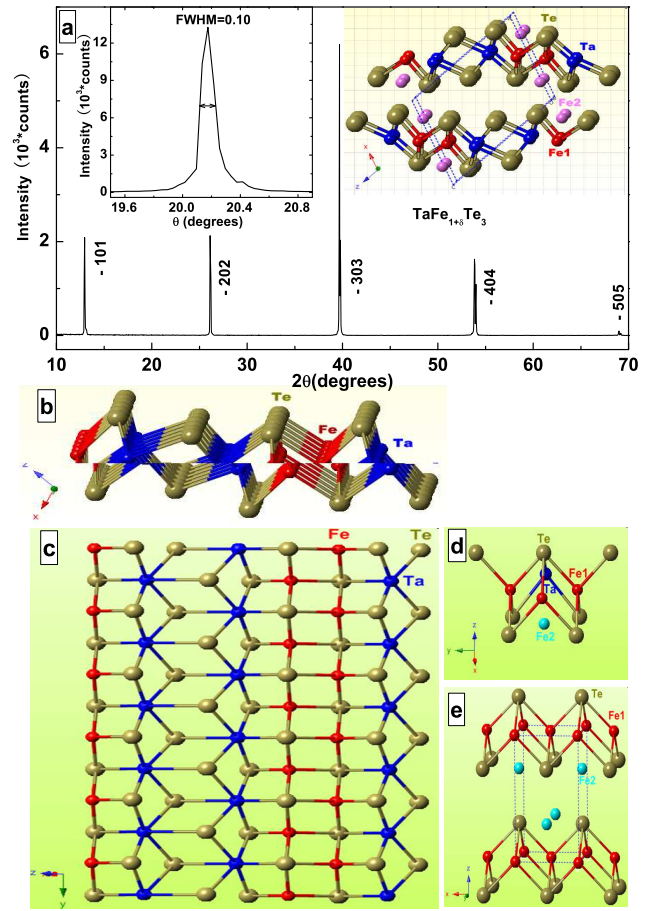}
\caption{(a) The single crystal X-ray diffraction pattern for
TaFe$_{1+y}$Te$_3$. The left inset shows the rocking curve at the
(-3 0 3) reflection. The crystal structure is shown in the right
inset; (b) A view of the TaFeTe$_3$ `` sandwiched" structure along
b-axis. There are two unique zigzag chains that are parallel to the
b-axis; (c) A view of the mixed metal network TaFeTe$_3$ along ($-l$
0 $l$); (d) The zigzag chain is made up of Fe centered edge-sharing
tetrahedra, similar to the FeTe$_4$ tetrahedra of anti-PbO type
Fe$_{1+y}$Te. The partial Fe2 atoms randomly occupy the interstitial
sites of the (Ta,Fe)Te layers; (E) The structure of anti-PbO type
Fe$_{1+y}$Te. The partial Fe2 atoms also randomly occupy the
interstitial sites of anti-PbO type FeTe layers.}
\end{figure}

Single crystals of TaFe$_{1+y}$Te$_3$ were grown by chemical vapor
transport method\cite{badding, liusx}. Ta (3N) powder, Fe (3N)
powder and Te (4N) powder were accurately weighed according to the
stoichiometric ratio of TaFe$_{1+y}$Te$_3$ (y = 0-0.25 ), then
thoroughly grounded and pressed into pellets. The 2 g mixture
pellets and 30 mg transport agent TeCl$_{4}$ were sealed in
evacuated 18 cm long $\times$ 15 mm diameter quartz tube. The sealed
tubes were placed in multi-zones tube furnace and slowly heated to
temperature with the hot end at 690 $\celsius$ and the cool end at
630 $\celsius$. After 150 hours, the furnace was shut off and cooled
down to room temperature. The long narrow crystals were obtained,
but in different temperature zones the samples had different
contents of interstitial Fe2. The sample preparation process except
for annealing was carried out in glove box in which high pure argon
atmosphere is filled. Single crystal X-ray diffraction (XRD) was
performed by MAC MXPAHF X-ray diffractometer (Japan) at room
temperature. Elemental analysis obtained by Energy Dispersive X-ray
Spectroscopy (EDX). Magnetic susceptibility measurement was
performed by a SQUID magnetometer (Quantum Design MPMS-XL7s).
Magnetoresistance (MR), Hall coefficient (R$_H$), heat capacity and
thermoelectric power (TEP) were measured by using Quantum Design
PPMS.

\begin{figure*}
\centering
\includegraphics[width=0.80\textwidth]{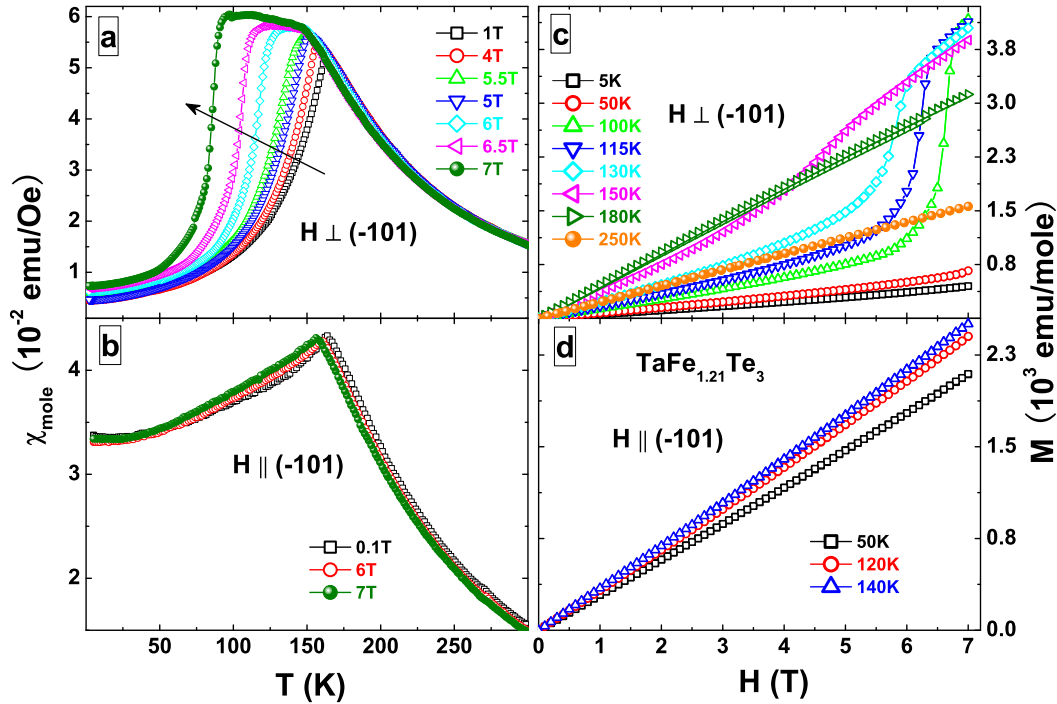}
\caption{Temperature dependence of susceptibility measured under
different fields  perpendicular to sample plane (a) or along sample
plane (b). Field dependence of magnetization $M$ at various
temperature with field perpendicular to sample plane (c) or along
sample plane (d).}
\end{figure*}

The structure of TaFe$_{1+y}$Te$_3$ is shown in the right inset of
Fig. 1a. The structure of TaFe$_{1+y}$Te$_3$ features a Ta-Fe bonded
network, and the mixed metal network lies between tellurium layers,
forming a FeTaTe$_3$ `` sandwich"\cite{badding, liusx, huang},
similar to that of anti-PbO type FeTe layer. It is stacked with
FeTaTe$_3$ `` sandwich" along ($-l$ 0 $l$), and crystallizes in
P21/m monoclinic symmetry with lattice constants a=7.4262 \AA,
b=3.6374 \AA, c=9.9925 \AA and $\beta$=109.166$^o$. According to
literatures\cite{badding, perez}, there are always partial excess
iron atoms, which randomly occupy the interstitial sites of
FeTaTe$_3$ layers\cite{perez}, similar to that in Fe$_{1+y}$Te. Fig.
1a shows XRD pattern of the platelet-shaped TaFe$_{1+y}$Te$_3$
single crystal at room temperature. Only ($-l$ 0 $l$) reflections
are observed in XRD pattern, indicating that the FeTaTe$_3$ ``
sandwich" plane parallel to the surface of the long piece-like
single crystal. The full width of half maximum (FWHM) in the rocking
curve of the (-3 0 3) reflection is 0.1$^o$, shown as the left inset
of Fig. 1a, suggesting it is a high quality single crystal. A view
of the FeTaTe$_3$ `` sandwich" structure along b-axis is shown in
Fig. 1b. There are two unique zigzag chains that are parallel to the
b-axis. The view nearly perpendicular to the Ta-Fe mixed network
layer is shown in Fig. 1c. One can easily see that one chain
consists of Ta-centered octahedra which shares Te-Ta edges. The
other chain is made up of Fe-centered edge-sharing tetrahedra. These
two zigzag chains alternately build up the Ta-Fe mixed networks.
From another point of view, the Ta-Fe mixed metal FeTaTe$_3$ layer
is made up of (Ta,Fe)Te tetrahedra Ta-Fe-Fe-Ta ribbons contacted by
sharing-edge Te-Te. The coordination environment of Fe1 in the
FeTaTe$_3$ `` sandwich" layer is the same as that of FeTe$_4$
tetrahedra in anti-PbO type FeTe layer. Additionally, there are
partial excess Fe2 atoms occupying randomly in square pyramidal
sites formed by five Te atoms. The (Ta,Fe)Te$_4$ tetrahedra
structure of FeTaTe$_3$ `` sandwich" layer and Fe$_{1+y}$Te
structure are shown in Fig. 1d and Fig. 1e, respectively.

Temperature dependence of susceptibility under different magnetic
fields perpendicular and parallel to the plane of single crystal are
shown in Fig.2a and Fig.2b, respectively. From the single crystal
XRD pattern, we know that the Ta-Fe mixed metal network is parallel
to the plane of single crystal. The susceptibility under low field
shows a sharp antiferromagnetic transition at T$_N$ $\sim$ 160 K,
which is lower than T$_N$=200 K of polycrystalline sample
TaFe$_{1.25}$Te$_3$. Elemental analysis of the single crystal with
T$_N$=160K obtained by EDX indicates that interstitial iron is about
0.21. This suggests that interstitial iron atoms Fe2 have strong
effect on the magnetism of the Fe zigzag chains. Less excess iron
will result in lower T$_N$. When external magnetic field is
perpendicular to the Ta-Fe mixed metal network, the AFM order
temperature T$_N$ is weakly suppressed from T$_N$=160 K at 0.1 T to
T$_N$=150 K at 7 T, and the more interesting thing is that high
field induces ferromagnetic behavior in susceptibility below T$_N$.
Below T$_N$ susceptibility firstly keeps constant with temperature
decreasing, then sharply drops at certain temperature T$_M$, which
depends on the value of external field H. At 160 K Fe1 of zigzag
chains form AFM order, and also induce that excess Fe2 of
interstitial sites to form antiferromagnetic order simultaneously
due to the direct coupling of Fe1 and Fe2. Below T$_N$ magnetic
moment of excess Fe2 tends to align along the direction of external
field H, and forms ferromagnetic order due to the coupling of Fe2
and external field. This is the reason why susceptibility keeps a
constant under higher vertical magnetic field below T$_N$. The
external field H$_{ext}$ and the inner field H$_{int}$ formed by Fe1
zigzag chains have a competition in tuning direction of magnetic
moment of Fe2 below T$_N$. The external field is in the ascendant in
tuning Fe2 to form FM alignment at high temperature, while the inner
field will be in the ascendant with temperature decreasing further,
which also induces the susceptibility to drop sharply at certain
temperature. The susceptibility of single crystal has strong
anisotropic magnetic properties at low field below T$_N$, suggesting
that the magnetic easy axis of Fe1 zigzag is along the [$-l$ 0 $l$].
Fig. 2c and 2d show field dependence of magnetization $M$ at various
temperature with field perpendicular and parallel to Ta-Fe mixed
network, respectively. When external magnetic field is perpendicular
to Fe1 zigzag of Ta-Fe mixed network, magnetization (\textbf{M})
increases linearly with the external field H$_{ext}$ at low field
and shows sharp FM-like increase at certain field H$_C$. But the
magnetic hysteresis is not observed in M-H curves at Fig. 2c and 2d.
It also arises from the result that the external field H competes
with the inner field formed by Fe1 zigzag chains for tuning the
direction of Fe2 spin.

\begin{figure}[t]
\includegraphics[width=0.50\textwidth]{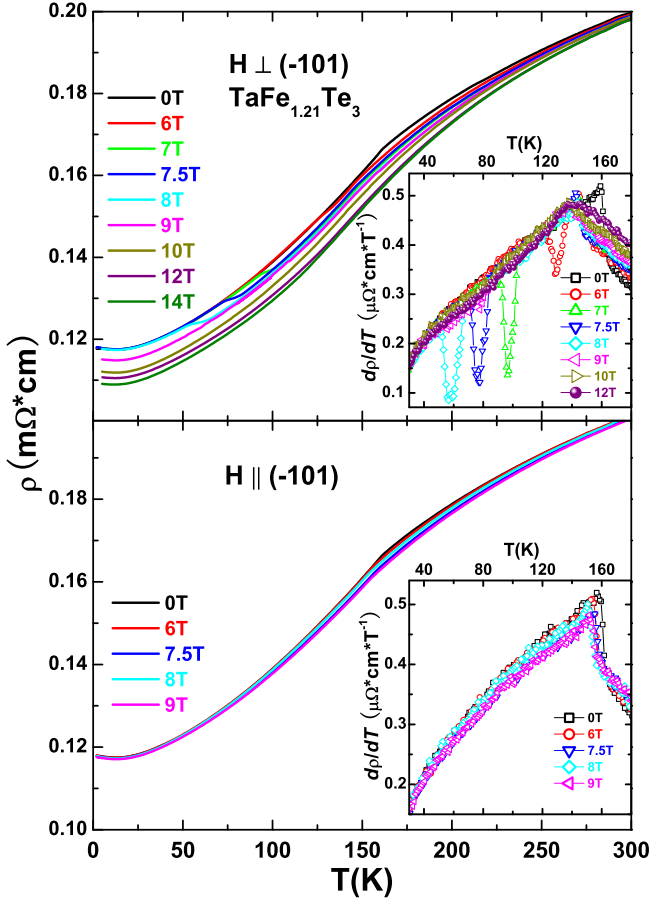}
\caption{Temperature dependence of resistivity under different
fields perpendicular to sample plane (up panel) or along sample
plane (down panel). Inset: Temperature dependence of their
derivative resistivity d$\rho$/d$T$.}
\end{figure}

\begin{figure}
\includegraphics[width=0.50\textwidth]{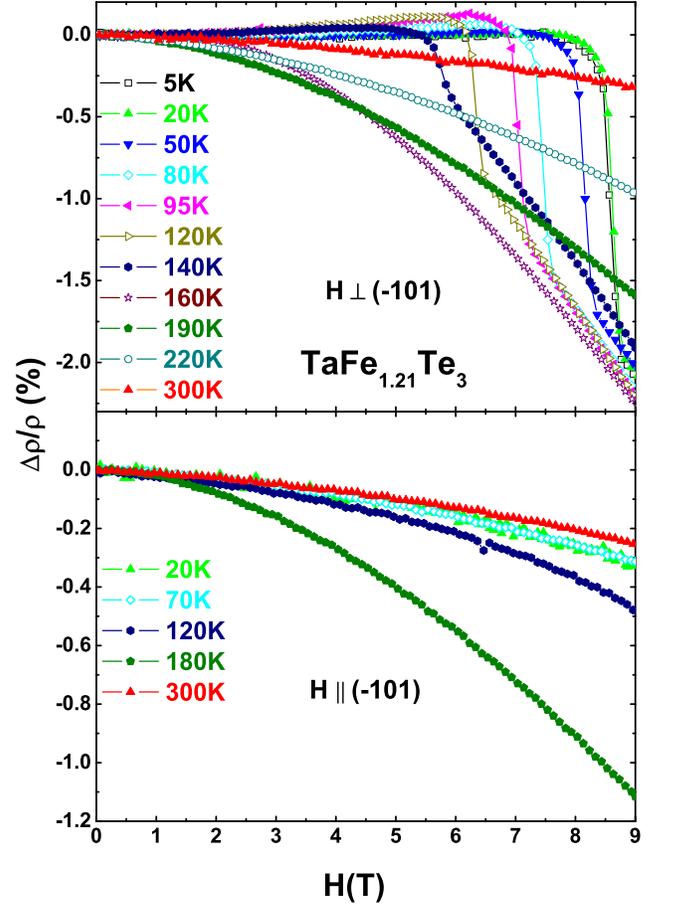}
\caption{Isothermal magnetoresistance at different temperatures with
field perpendicular to sample plane (up panel) or along sample plane
(down panel).}
\end{figure}

Figure 3 shows temperature dependence of resistivity in various
magnetic field H. The resistivity shows a metal behavior and has an
abnormal transition at T$_{S1}$$\sim$160 K under low field,
corresponding to the drop of magnetic susceptibility. When external
magnetic field is parallel to the Ta-Fe mixed metal network, the
abnormal transition temperature T$_{S1}$ has almost no change at
different fields. However, under high field perpendicular to the
single crystal plane T$_{S1}$ is weakly suppressed to low
temperature (T$_{S1}$=150 K for H $>$ 6 T) and the resistivity has
another abnormal transition simultaneously at certain low
temperature T$_{S2}$, which depends on the magnitude of external
field H. These abnormal behaviors of resistivity can be easily seen
in these inters of Fig.3. The T$_{S1}$ and T$_{S2}$ in resistivity
correspond with the AFM transition temperature T$_N$ of Fe1 and the
metamagnetic transition temperature T$_M$ of Fe2 in magnetic
susceptibility, respectively.

\begin{figure}
\includegraphics[width=0.50\textwidth]{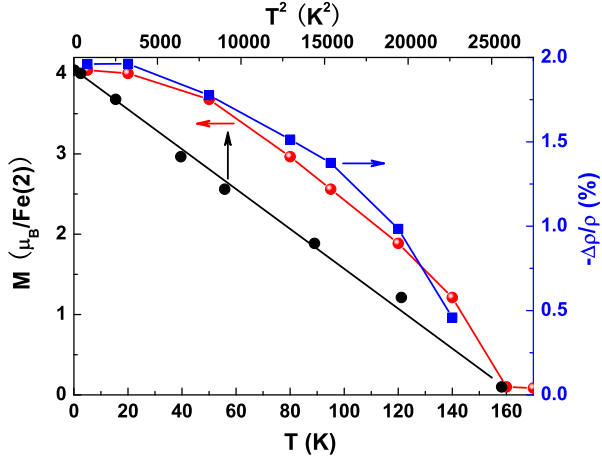}
\caption{Magnetization of Fe2 \emph{\textbf{M}} vs T (Light Y) and
MR $\Delta \rho(T)/\rho(T)$ vs T (Left Y). \emph{\textbf{M}} vs
T$^2$ shows that $\Delta M(T)/M(0) \propto T^2$ up to 160 K.}
\end{figure}

To confirm the origin of the anomalous in resistivity, isothermal
magnetoresistance (MR) at different temperature is shown in Fig.4.
One can easily see that they have negative MR effect under parallel
field for all temperature and under vertical field for higher than
T$_N$. For vertical field, they have small positive MR below the
certain field H$_C$, and it shows a sharp negative MR around H$_C$,
which should be ascribed to spin-flop of Fe2 at certain field H$_C$.
The largest negative MR effect appearing at a little higher
temperature than T$_N$ suggests that there is strong magnetic
fluctuation around T$_N$. The temperature dependence of normal Hall
coefficient R$_H$ also supports this point of view.

Figure 5 shows the `` spontaneous" magnetization \emph{\textbf{M}}
of Fe2, which is inferred from the jump magnitudes $\Delta M$ in the
isothermal magnetization. The saturation magnetization at 5 K is 4
$\pm $ 0.2 $\mu_{\textrm{B}}$ per Fe2, which is the same with local
moment Fe (4 $\mu_{\textrm{B}}$/Fe) in Fe$_{1/4}$TaS$_2$. The
transition metal intercalated dichalcogenide Fe$_{1/4}$TaS$_2$ has a
ferromagnetic transition at T$_C$ $\sim$ 160 K, in which the
spontaneous magnetization of Fe is strongly pinned perpendicular to
the TaS$_2$ layers by a very large anisotropy field below
T$_C$\cite{ong}. Neutron scattering experiments reveal that the Fe
has a total moment 2.25(8) $\mu_{\textrm{B}}$/Fe in
Fe$_{1+y}$Te\cite{li}. Since the moments of the partial Fe2 ions are
randomly distributed in the interstitial sites of FeTe layers, it is
difficult to estimate the moment sizes of excess Fe2 by using
conventional neutron diffraction\cite{Bao4, li}. However, the
theoretical calculation suggests that the excess Fe2 has very strong
magnetism with high local moment\cite{zhang}. We find that
magnetization of excess Fe2 has a near perfect T$^2$ dependence
below T$_N$, in agreement with normal ferromagnetic
metals\cite{zeng}, as shown in Fig. 5. The spin-flop of Fe2 tuned by
field induces a sharp negative MR ($\Delta \rho(T)/\rho(T)$) around
H$_C$, as shown in Fig. 5. $\Delta \rho(T)/\rho(T)$ and
magnetization \emph{\textbf{M}} of Fe2 have the same dependence of
temperature. $\frac{\Delta \rho/\rho}{M_{Fe2}}$ is about 0.45
\%/$\mu_{\textrm{B}}$(Fe2). We use the Curie-Weiss expression $\chi
= \chi_o + C/(T +\Theta)$ to fit the susceptibility data from 330 to
400 K, where $C$ is the Curie constant, $\Theta$ the Weiss
temperature and $\chi_o$ constant. The total effective moment of 3.9
$\mu_{\textrm{B}}$ per iron atom is obtained from the Curie
constant. It is a little larger than the result 3.7
$\mu_{\textrm{B}}$ in previous literature, in which the
susceptibility of polycrystalline sample is fitted by the
Curie-Weiss expression from 450 to 1000 K\cite{badding}. From MR and
the following Hall coefficient result, we know that there is very
strong magnetic fluctuation or magnetic correlation between iron
atoms above T$_N$. The temperature range of susceptibility fitted
from 330 to 400 K is too low and brings some deviation of total
effective moment inferred. In spite of the small deviation, the
magnetic moment of the Fe1 (3.7 $\pm0.2$ $\mu_{\textrm{B}}$) in
Ta-Fe-Fe-Ta ribbons of the TaFeTe$_3$ ``sandwich" is almost two
times larger than that of Fe1 (2.25 $\mu_{\textrm{B}}$) in the
anti-PbO type FeTe layers.

\begin{figure}
\includegraphics[width=0.50\textwidth]{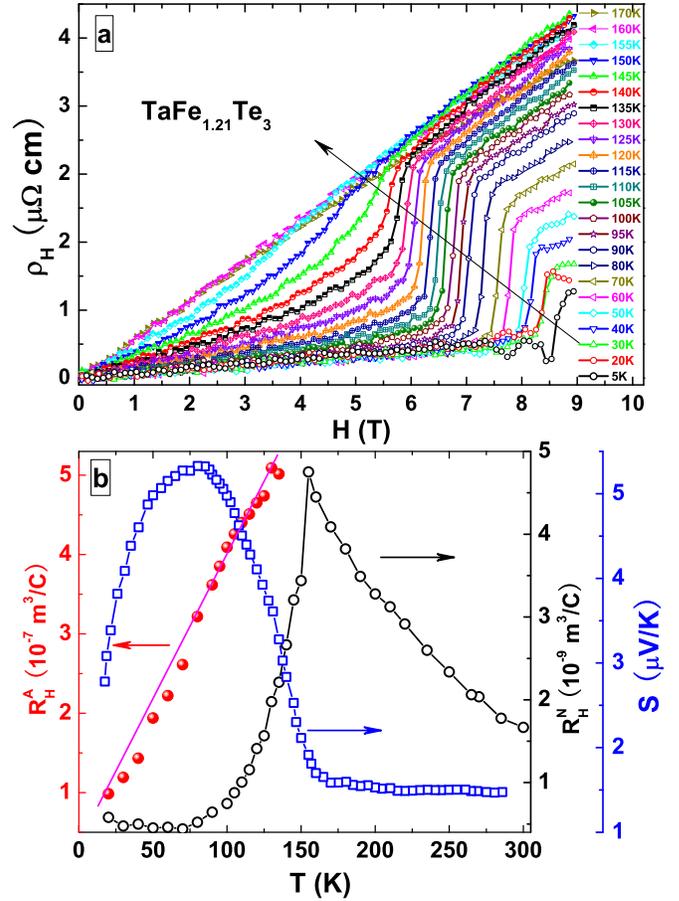}
\caption{(a) Field dependent Hall resistivity $\rho_{H}$ at various
temperatures are obtained by using
$[\rho_{xy}(+H)-\rho_{xy}(-H)]/2$, where $\rho_{xy}(\pm H)$ is
$\rho_{xy}$ under positive or negative magnetic field. (b) The
temperature dependence of R$^A_H$, R$^N_H$ and thermoelectric power.
The anomalous Hall effect coefficient R$^A_H$ is inferred from the
jump in $\rho_{H}$ at H$_C$, while the normal Hall effect
coefficient R$^N_H$ is inferred from the \emph{H}-linear portions of
$\rho_{H}$ below the ``coercive" field H$_C$. }
\end{figure}

It is well known that Hall effect arises from two parts-normal Hall
effect and anomalous Hall effect in ferromagnetic metals, in which
anomalous Hall resistivity is proportional to the magnetization M.
Empirically, one finds Hall resistivity $\rho_H = \rho_{OH} +
\rho_{AH} = R^N_H + R^A_H4\pi M$, where $\rho_{OH}$ is the normal
Hall resistivity due to the Lorentz force in a perpendicular
magnetic field \emph{\textbf{B}}, $\rho_{AH}$ the anomalous Hall
resistivity, $R^N_H$ the normal Hall coefficient, and $R^A_H$ the
anomalous Hall coefficient\cite{nagaosa}. To confirm the origin of
the AFM transition of Fe1 zigzag chains and field inducing FM
transition of Fe2, the transverse resistivity $\rho_{xy}$ is
measured by sweeping field from -9 T to +9 T at various temperature,
and the accurate Hall resistivity $\rho_H$ is obtained, as shown in
Fig. 6a, by using $[\rho_{xy}(+H)-\rho_{xy}(-H)]/2$, where
$\rho_{xy}(\pm H)$ is $\rho_{xy}$ under positive or negative
magnetic field. Similar to the isothermal MR at various temperature,
$\rho_H$ also shows a steep rise at certain field H$_C$ below T$_N$,
which arises from the jump magnitudes of magnetization \textbf{M}
due to the spin-flop of excess Fe2 induced by external field
\textbf{H}. The normal Hall coefficient R$_H$ is obtained from these
H-linear term of $\rho_H$ below the `` coercive" field H$_C$.
R$^N_H$ dependence of temperature is shown in Fig. 5. The R$^N_H$ is
positive, indicating the carrier is hole. Above T$_N$, R$^N_H$
decreases distinctly with temperature increasing and almost has a
linear dependence of temperature. It may arise from the strong
magnetic fluctuation above T$_N$ in this system. Empirically, the
change of R$^N_H$ is weakly dependent on temperature above T$_C$ in
general ferromagnetic metal. R$^N_H$ shows a pronounced dip at
T$_N$. It should be ascribed to that the magnetic fluctuation is
suppressed completely and the transport lifetime $\tau$ has a strong
change around the fermi surface below T$_N$. The Hall number density
$n_H=1/eR^N_H$ varies from the minimal value 1.3$\times$10$^{21}$
cm$^{-3}$ at 155 K to 1.4 $\times$10$^{22}$ cm$^{-3}$ at 75 K. The
normal Hall coefficient R$^N_H$ is the same to that of
Fe$_{1+y}$Te(R$_H$ $\sim$ 10$^{-9} m^3$/C)\cite{liutj}. The
thermoelectric power (TEP) is positive and has the same sign of the
Hall coefficient. As shown in Fig. 6b, TEP shows a weak temperature
dependence above T$_N \sim$ 160 K, but it has a pronounced rise
below T$_N$ and arrives at the maximum (5.2 $\mu$V/K) around 75 K.
The resistivity, susceptibility, R$^N_H$ and TEP show anomalous
behaviors below T$_N$, which are very similar to that in metal Cr
around SDW-type AFM transition\cite{fawcett}. In addition, the
structure of TaFe$_{1+y}$Te$_3$ features a Fe1 zigzag chains along
b-axis. It is well known that many low dimensional materials have
SDW and charge density wave (CDW) instability at low
temperature\cite{gruner2,gruner3}. The behaviors of these physical
properties at T$_N$ suggest that the transition should be a SDW-type
AFM transition. The anomalous Hall coefficient R$^A_S$ below T$_N$
is inferred from the ratio of the jump magnitudes $\Delta M$ and
$\delta \rho_H$ around H$_C$. R$^A_S$=$\Delta\rho_{H}/4\pi M$ is
also plotted in Fig .6b. R$^A_S$ decreases linearly with temperature
below T$_N$.

\begin{figure}[h]
\includegraphics[width=0.50\textwidth]{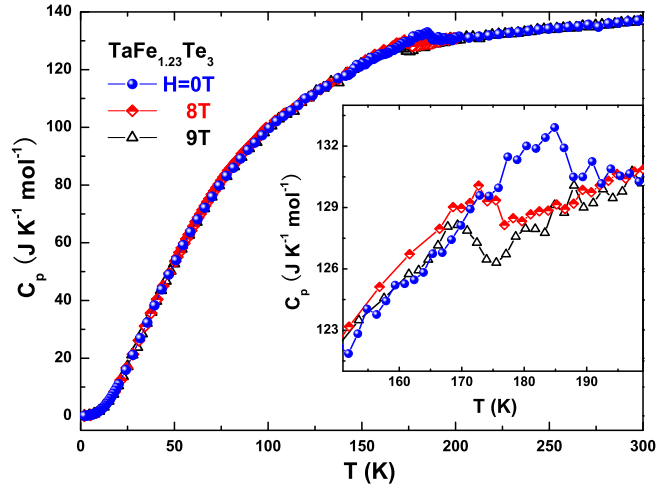}
\caption{Temperature dependence of specific heat for
TaFe$_{1.23}$Te$_3$ under different fields perpendicular to sample
plane. }
\end{figure}

In order to confirm the AFM transition, the heat capacity was
measured by a relaxation-time method with a Quantum Design PPMS. One
can clearly see a pronounced anomaly peak of C$_p$(T) at $~$185 K
for zero field in Fig.7. The temperature is different from the T$_N$
inferred by susceptibility due to different samples with different
magnitude of excess Fe. The sample measured specific heat has 0.23
interstitial Fe2 confirmed by EDX. The temperature of specific heat
peak is shifted from 185 K to 170 K at 9 T in excellent agreement
with the previous susceptibility and MR under high field. They
consistently confirm that high magnetic field suppress the AFM order
of Fe1 zigzag chains distinctly. In the low-temperature region the
specific heat is of the form C$_p$ = $\gamma$T + $\beta$T$^3$. The
Debye temperature can be estimated from the equation $\beta$ =
(12$\pi$$^4$Nk$_B$)/(5$\Theta$$^3$$_D$), where N is the number of
atoms per formula unit. From the plot of C$_p$/T vs.T$^2$ data
between 2 and 14 K, we can estimate a Sommerfeld coefficient
$\gamma$ = 25.86 mJ K$^{-2}$ mol$^{-1}$, $\beta$ = 1.496 mJ K$^{-4}$
mol$^{-1}$ and $\Theta$$_D$ = 189 K for TaFe$_{1.23}$Te$_3$. The
electron specific heat coefficient $\gamma$ is close to that of
Fe$_{1+y}$Te ($\gamma$=27 mJ K$^{-2}$ mol$^{-1}$)\cite{liutj}. The
above resistivity and normal Hall coefficient R$^N_H$ also show that
Fe$_{1+y}$Te and TaFe$_{1+y}$Te$_3$ have the same order of
magnitude. It suggests that Fe$_{1+y}$Te and TaFe$_{1+y}$Te$_3$ have
almost the same density of states near the Fermi energy level.

\begin{figure}[h]
\includegraphics[width=0.45\textwidth]{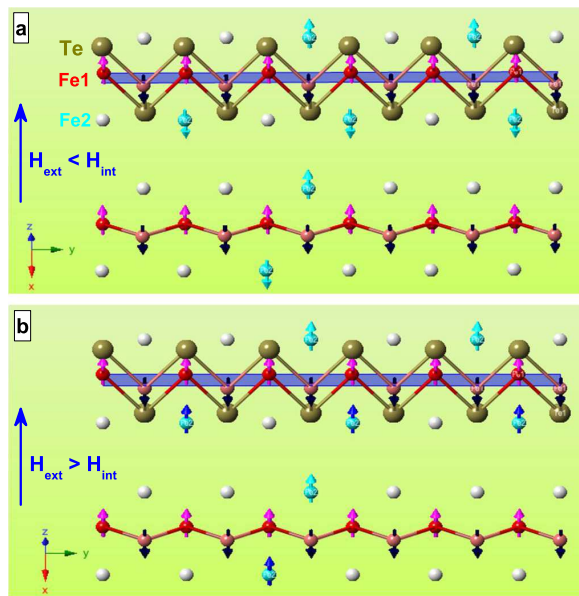}
\caption{The spin model of excess Fe and Fe zigzag chain with
external magnetic field H along ($-l$ 0 $l$): (a) 0 $\leq$ H$_{ext}
< $H$_{int}$; (b) H$_{ext} > $H$_{int}$.}
\end{figure}

Based on the above results of susceptibility, MR and Hall effect,
possible magnetic structures for the spins of Fe1 and Fe2 are
proposed shown in Fig. 8a and 8b. Below T$_N$, Fe1 atoms of zigzag
chains form antiferromagnetic alignment. The magnetic easy axis of
Fe1 should be perpendicular to the Ta-Fe mixed metal network layers
with a very large anisotropy energy. Since excess Fe2 directly
couples with Fe1 of zigzag chains, the random excess Fe2 forms
ferromagnetic alignment with nearest two Fe1 due to the inner field
H$_{int}$ induced by Fe1 in zigzag chain below T$_N$. Furthermore,
the inner field and the coupling energy between Fe2 and Fe1 are both
enhanced with temperature decreasing. On the other hand, the
external field $H_{ext}$ has very weak suppression on the AFM
transition of Fe1 in zigzag chains, but strongly tunes the direction
of Fe2 spin as long as $H_{ext}$ $>$  $H_{int}$. Since the especial
crystal structure and magnetic structure shown in Fig. 8, the inner
field will induce the excess Fe2 up and down among Fe1 zigzag chains
to form AFM alignment between Fe2(up) and Fe2(down). That is why the
whole Fe form AFM alignment below T$_N$ under low field. However,
the external field prefers the all excess Fe2 to be parallel with
each other along external field. When the external field overcomes
the inner field with temperature increasing, it will tune Fe2 atoms
which are antiparallel to external field to reverse their spins.
This causes the fact that susceptibility sharply increases at the
certain temperature T$_C$ or certain field H$_C$, as shown in Fig.
2.

In summary, we systematically study the AFM order of Fe1 zigzag
chains and spin-flop of excess Fe2 under high magnetic field H
through the susceptibility, MR, Hall effect and specific heat
measurements in high-quality single crystal TaFe$_{1+y}$Te$_3$.
These properties suggest that the high temperature AFM transition of
the TaFeTe$_3$ layers should be a SDW-type AFM order. Below T$_N$
Fe1 antiferromangetic chains will induce a inner magnetic field
\textbf{H$_{int}$} to excess Fe2 and lead Fe2 to form an AFM
alignment, in which the magnetic coupling strength between Fe1 and
Fe2 is enhanced by decreasing temperature. On the other hand, the
external magnetic field \textbf{H$_{ext}$} competes with the inner
magnetic field \textbf{H$_{int}$} induced by AFM order of Fe1 zigzag
chains and inclines to tune excess Fe2 to form FM alignment along
\textbf{H$_{ext}$}. The excess Fe2 has a spin-flop at the "coercive"
field H$_C$, where \textbf{H$_{ext}$} can overcome the
\textbf{H$_{int}$}. Based on spin-flop of Fe2, the local moment of
Fe2(4 $\mu_{\textrm{B}}$/Fe) can be obtained from $\Delta$\textbf{M}
around H$_C$ in M-H curves. The possible magnetic structure of
TaFe$_{1+y}$Te$_3$ is also proposed. The properties related to the
spin-flop of Fe2 supply a good opportunity to study the coupling
between Fe1 and Fe2 in these TaFe$_{1+y}$Te$_3$ or Fe$_{1+y}$Te
compounds with excess Fe2.

{\bf Acknowledgment:} This work is supported by the Natural Science
Foundation of China and by the Ministry of Science and Technology of
China (973 project No:2006CB601001) and by Natural Basic Research
Program of China (2006CB922005).


\begin{thebibliography}{\clearpage}

\bibitem{Kamihara}
Y. Kamihara, T. Watanabe, M. Hirano, and H. Hosono, J. Am. Chem.
Soc. {\bf 130}, 3296 (2008).

\bibitem{xhchen}
X. H. Chen, T. Wu, G. Wu, R. H. Liu, H. Chen, D. F. Fang, Nature
{\bf 354}, 761-762(2008).

\bibitem{ren3}
Z. A. Ren, G. C. Che, X. L. Dong, J. Yang, W. Lu, W. Yi, X. L. Shen,
Z. C. Li, L. L. Sun, F. Zhou, et al., Europhysics Letters, {\bf 83},
17002(2008).

\bibitem{hsu}
F.-C. Hsu et al., Proc. Natl. Acad. Sci. {\bf105}, 14262(2008).

\bibitem{yeh}
K.-W. Yeh, T.-W. Huang, Y.-L. Huang, T.-K. Chen, F.-C. Hsu, P. M.
Wu, Y.-C. Lee, Y.-Y. Chu, C.-L. Chen, J.-Y. Luo, D. C. Yan, and M.
K. Wu, Europhys. Lett. {\bf84}, 37002(2008).

\bibitem{fang}
M. H. Fang, H. M. Pham, B. Qian, T. J. Liu, E. K. Vehstedt, Y. Liu,
L. Spinu, and Z. Q. Mao, Phys. Rev. B {\bf78}, 224503(2008).

\bibitem{mizuguchi}
Y. Mizuguchi, F. Tomioka, S. Tsuda, T. Yamaguchi, and Y. Ta- kano,
Appl. Phys. Lett. {\bf93}, 152505(2008).

\bibitem{medvedev}
S. Medvedev, T. M. McQueen, I. A. Troyan, T. Palasyuk, M. I.
Eremets, R. J. Cava, S. Naghavi, F. Casper, V. Ksenofontov, G.
Wortmann, C. Felser, Nature materials, {\bf8}, 630(2009).

\bibitem{xlchen}
J. Guo, S. Jin, G. Wang, S. Wang, K. Zhu, T. Zhou, M. He and X.
Chen, Phys. Rev. B {\bf 82}, 180520 (2010).

\bibitem{fang1}
Minghu Fang, Hangdong Wang, Chiheng Dong, Zujuan Li, Chunmu Feng,
Jian Chen, H.Q. Yuan, EPL {\bf 94}, 27009(2011) .

\bibitem{Mizuguchi} 
Yoshikazu Mizuguchi, Hiroyuki Takeya, Yasuna Kawasaki, Toshinori
Ozaki, Shunsuke Tsuda, Takahide Yamaguchi and Yoshihiko Takano,
Appl. Phys. Lett. {\bf98}, 042511 (2011).

{\bibitem{Wang}
 A. F. Wang, J. J. Ying, Y. J. Yan, R. H. Liu, X. G.
Luo, Z. Y. Li, X. F. Wang, M. Zhang, G. J. Ye, P. Cheng, Z. J.
Xiang, X. H. Chen, Phys. Rev. B {\bf83}, 060512(R)(2011).


\bibitem{Bendele}
M. Bendele et al., Phys. Rev. Lett. \textbf{104}, 087003(2010).

\bibitem{Margadonna}
S. Margadonna et al., Phys. Rev. B \textbf{80}, 064506(2009).


\bibitem{sher}
Z. Shermadini, A. Krzton-Maziopa, M. Bendele, R. Khasanov, H.
Luetkens, K. Conder, E. Pomjakushina, S. Weyeneth, V. Pomjakushin,
O. Bossen, A. Amato, Phys. Rev. Lett. {\bf106}, 117602(2011).

\bibitem{BaoWW} 
Wei Bao, Q. Huang, G. F. Chen, M. A. Green, D. M. Wang, J. B. He, X.
Q. Wang, Y. Qiu, Chinese Phys. Lett. {\bf28}, 086104(2011).

\bibitem{Vyu} 
V. Yu. Pomjakushin, D. V. Sheptyakov, E. V. Pomjakushina, A.
Krzton-Maziopa, K. Conder, D. Chernyshov, V. Svitlyk, Z. Shermadini
, Physical Review B {\bf83}, 144410(2011).

\bibitem{lrh}
R. H. Liu, X. G. Luo, M. Zhang, A. F. Wang, J. J. Ying, X. F. Wang,
Y. J. Yan, Z. J. Xiang, P. Cheng, G. J. Ye, Z. Y. Li and X. H. Chen,
EPL, {\bf94}, 27008(2011).

\bibitem{fredrik}
Fredrik Gr${\o}$nvold, Haakon Haraldsen and John Vihovde, Acta
Chemica Scandinavics \textbf{8}, 1927-1942(1954).

\bibitem{fruchart}
D. Fruchart, P. Convert, P. Wolfers, R. Madar, J. P. Senateur, and
R. Fruchart, Mater. Res. Bull. \textbf{10}, 169(1975).

\bibitem{hisao}
Katsuraki Hisao, Achiwa Norio, Journal of the Physical Society of
Japan, \textbf{21}, 2238(1966).

\bibitem{ccruz}
C. Cruz, Q. Huang, J. W. Lynn, J. Li, W. Ratcliff, J. L. Zarestky,
H. A. Mook, G. F. Chen, J. L. Luo, N. L.Wang, et al., Nature {\bf
453}, 899 (2008).

\bibitem{huang}
Q. Huang, Y. Qiu, Wei Bao, M. A. Green, J.W. Lynn, Y. C. Gasparovic,
T. Wu, G. Wu, and X. H. Chen,  Phys. Rev. Lett  {\bf 101}, 257003
(2008).

\bibitem{Bao4}
W. Bao, Y. Qiu, Q. Huang, M.A. Green, P. Zajdel, M.R. Fitzsimmons,
M. Zhernenkov, M. Fang, B. Qian, E.K. Vehstedt, J. Yang, H.M. Pham,
L. Spinu, Z.Q. Mao, Phys. Rev. Lett. \textbf{102}, 247001 (2009).

\bibitem{li}
Shiliang Li, Clarina de la Cruz, Q. Huang, Y. Chen, J. W. Lynn,
Jiangping Hu, Yi-Lin Huang, Fong-Chi Hsu, Kuo-Wei Yeh, Maw-Kuen Wu,
and Pengcheng Dai, Phys. Rev. B \textbf{79}, 054503(2009).


\bibitem{zhang}
Lijun Zhang, D. J. Singh and M. H. Du, Phys. Rev. B{\bf 79},
012506(2009).

\bibitem{liutj}
T. J. Liu, X. Ke, B. Qian, J. Hu, D. Fobes, E. K. Vehstedt, H. Pham,
J. H. Yang, M. H. Fang, L. Spinu, P. Schiffer, Y. Liu, and Z. Q.
Mao, Phys. Rev. B {\bf80}, 174509(2009).

\bibitem{badding}
M. E. Badding, J. Li, F. J. Disalvo, W. Zhou and P. P. Edwards,
Journal of solid state chemistry {\bf100}, 313-324(1992).

\bibitem{liusx}
Shi-xiong Liu, Guan-liang Cai and Jin-ling Huang, Acta Cryst. C {\bf
49}, 4-7 (1993).

\bibitem{perez}
C. P\'{e}rez Vicente, M. Womes, J. C. Jumas, L. S\'{a}nchez and J.
L. Tirado, J. Phys. Chem. B {\bf102}, 8712-8718(1998)

\bibitem{huang}
Jinling Huang, Science in China, {\bf43}, 337-347(2000).

\bibitem{nagaosa} 
Naoto Nagaosa, Jairo Sinova, Shigeki Onoda, A. H. MacDonald, N. P.
Ong, Rev. Mod. Phys. {\bf82}, 1539(2010).

\bibitem{ong} 
J. G. Checkelsky, Minhyea Lee, E. Morosan, R. J. Cava and N. P. Ong,
Phys. Rev. B {\bf77}, 014433(2008).

\bibitem{zeng} 
Changgan Zeng, Yugui Yao, Qian Niu, Hanno H. Weitering, Phys. Rev.
Lett. {\bf96}, 037204(2006).

\bibitem{fawcett}
E. Fawcett, Rev. Mod. Phys. \textbf{60}, 209 (1988).

\bibitem{gruner2}
G. Gr\"{u}ner, Rev. Mod. Phys.\textbf{66}, 1-24 (1994).

\bibitem{gruner3}
George Gr\"{u}ner: Density waves in solids, Addison-Wesley, 1994

}

\end{thebibliography}
\end{document}